\documentclass[aps,physrev,twocolumn,superscriptaddress,10pt,floatfix]{revtex4-2}

\usepackage{amsmath}
\usepackage{graphicx}
\usepackage{color}

\usepackage{hyperref}
\hypersetup{colorlinks=true,linkcolor=red,anchorcolor=blue,citecolor=blue, filecolor=blue,urlcolor=red,bookmarksnumbered=true,
	pdfview=FitB
}

\begin{document}

\title{Exclusive $J/\psi$ photoproduction in photon-proton diffractive scattering: A light-front Hamiltonian approach}

\author{Xiaoyi Wu}
\email{wuxiaoyi@impcas.ac.cn}
\affiliation{Institute of Modern Physics, Chinese Academy of Sciences, Lanzhou 730000, China}
\affiliation{School of Nuclear Science and Technology, University of Chinese Academy of Sciences, Beijing 100049, China}
\affiliation{CAS Key Laboratory of High Precision Nuclear Spectroscopy, Institute of Modern Physics, Chinese Academy of Sciences, Lanzhou 730000, China}

\author{Zhi Hu}
\email{huzhi@post.kek.jp}
\affiliation{High Energy Accelerator Research Organization (KEK), Ibaraki 305-0801, Japan}

\author{Chandan Mondal}
\email{mondal@impcas.ac.cn}
\affiliation{Institute of Modern Physics, Chinese Academy of Sciences, Lanzhou 730000, China}
\affiliation{School of Nuclear Science and Technology, University of Chinese Academy of Sciences, Beijing 100049, China}
\affiliation{CAS Key Laboratory of High Precision Nuclear Spectroscopy, Institute of Modern Physics, Chinese Academy of Sciences, Lanzhou 730000, China}

\author{Jiangshan Lan}
\email{jiangshanlan@impcas.ac.cn}
\affiliation{Institute of Modern Physics, Chinese Academy of Sciences, Lanzhou 730000, China}
\affiliation{School of Nuclear Science and Technology, University of Chinese Academy of Sciences, Beijing 100049, China}
\affiliation{CAS Key Laboratory of High Precision Nuclear Spectroscopy, Institute of Modern Physics, Chinese Academy of Sciences, Lanzhou 730000, China}

\author{Siqi Xu}
\email{xsq234@impcas.ac.cn}
\affiliation{Institute of Modern Physics, Chinese Academy of Sciences, Lanzhou 730000, China}
\affiliation{School of Nuclear Science and Technology, University of Chinese Academy of Sciences, Beijing 100049, China}
\affiliation{Department of Physics and Astronomy, Iowa State University, Ames, IA 50011, U.S.A.}

\author{Jiatong Wu}
\email{wujt@impcas.ac.cn}
\affiliation{Institute of Modern Physics, Chinese Academy of Sciences, Lanzhou 730000, China}
\affiliation{School of Nuclear Science and Technology, University of Chinese Academy of Sciences, Beijing 100049, China}
\affiliation{CAS Key Laboratory of High Precision Nuclear Spectroscopy, Institute of Modern Physics, Chinese Academy of Sciences, Lanzhou 730000, China}

\author{Xingbo Zhao}
\email{xbzhao@impcas.ac.cn}
\affiliation{Institute of Modern Physics, Chinese Academy of Sciences, Lanzhou 730000, China}
\affiliation{School of Nuclear Science and Technology, University of Chinese Academy of Sciences, Beijing 100049, China}
\affiliation{CAS Key Laboratory of High Precision Nuclear Spectroscopy, Institute of Modern Physics, Chinese Academy of Sciences, Lanzhou 730000, China}

\author{James P. Vary}
\email{jvary@iastate.edu}
\affiliation{Department of Physics and Astronomy, Iowa State University, Ames, IA 50011, U.S.A.}

\collaboration{BLFQ Collaboration}
\noaffiliation

\date{\today}

\begin{abstract}
We investigate the cross-section for exclusive $J/\psi$ production in photon-proton diffractive scattering within the Basis Light-Front Quantization (BLFQ) framework. The leading-order contribution to this process is well approximated by the charge conjugation-even two-gluon (``pomeron") exchange mechanism in the dipole model, which factorizes the total amplitude into the dipole scattering amplitude and the convolution of the $J/\psi$ and photon light-front wave functions (LFWFs). We express the dipole scattering amplitude as the matrix element of gluon field operators inserted between proton states, with the element being sensitive to the proton LFWFs and the Bjorken scaling variable, $x$. The proton and $J/\psi$ LFWFs are obtained by diagonalizing their respective light-front Hamiltonians within the BLFQ approach, while the virtual photon LFWFs are employed from perturbative QCD. Our results provide initial conditions for the Balitsky-Kovchegov (BK) equation, which can be used to probe the proton structure at smaller Bjorken scales. This work offers valuable theoretical insights for future electron-ion collider experiments.
\end{abstract}


\maketitle

\section{Introduction\label{Sec1}}
The exploration of quantum chromodynamics (QCD) at small Bjorken scaling variable $x$ is a central frontier in understanding the high-energy limit of strong interactions. In this regime, gluon distributions inside hadrons become overwhelmingly dominant. Investigations in this field are crucial for confirming the existence of the Color Glass Condensate (CGC)~\cite{mclerran1994gluon}, a novel state that emerges from extreme gluon densities. Key processes in electron-proton  collisions such as deep inelastic scattering (DIS), deeply virtual Compton scattering (DVCS), and exclusive vector meson production (VMP) serve as probes of this saturated gluonic matter~\cite{nikolaev1991colour}, providing essential insight into the transition from linear evolution to high-density QCD. Moreover, small-$x$ physics is central to resolving the enduring mysteries of the proton’s mass generation and spin composition, as the high-density gluons with low-momentum fraction are responsible for the majority of the mass and a significant contribution to the spin. It is also regarded as an important testing ground for strong interaction dynamics in the nonperturbative regime of QCD.

Experimentally, this field has advanced through hadron-electron interactions at the Hadron Electron Ring Accelerator (HERA)~\cite{H1:2018flt} facility, as well as ultra-peripheral collisions in proton-proton, proton-nucleus, and nucleus-nucleus interactions at the Large Hadron Collider (LHC)~\cite{CMS:2016itn,ALICE:2015mzu} and the Relativistic Heavy Ion Collider (RHIC)~\cite{STAR:2023vvb} facilities. Additionally, the Electron-Ion Collider (EIC)~\cite{AbdulKhalek:2021gbh} and the Electron-ion collider in China (EicC) are currently in the planning stages~\cite{Anderle:2021wcy}.

One of the key processes measured in these experiments is exclusive $J/\psi$ production in photon-proton diffractive scattering~\cite{H1:2005dtp,ZEUS:2004yeh,LHCb:2018rcm,STAR:2023vvb}. The VMP process is known to be highly sensitive to hadronic target wave function fluctuations. The $J/\psi$ channel, with its enhanced experimental accessibility compared to lighter vector mesons like the $\phi$ and $\rho$, serves as a powerful probe of the proton's internal gluonic structure at small $x$. The leading-order contribution to this process can be well approximated using the dipole model, which factorizes the scattering amplitude of the process $\gamma^* p \to V p$ (here, $V$ denotes a vector meson) into the dipole scattering amplitude and the convolution of the $J/\psi$ and photon light-front wave functions (LFWFs)~\cite{kowalski2006exclusive}. While the photon’s LFWFs in the valence Fock sector are known exactly in perturbative QCD, determining the LFWFs of vector mesons remains challenging due to their nonperturbative nature. Similarly, the dipole scattering amplitude is closely related to the proton’s LFWFs, which are also difficult to compute from first principles.

A novel theoretical approach has been proposed in Ref.~\cite{Dumitru:2019qec}, utilizing ``harmonic oscillator" and ``power law" model wave functions for the proton~\cite{brodsky1994wavefunction} as non-perturbative inputs to compute the dipole scattering amplitude at moderate $x$ under the eikonal approximation. In addition to the model employed in Ref.~\cite{Dumitru:2019qec}, numerous other phenomenological models and theoretical frameworks have been developed to obtain the LFWFs of the $J/\psi$ or proton. These include approaches such as the Dyson-Schwinger and Bethe-Salpeter frameworks~\cite{Shi:2021taf}, the constituent quark model~\cite{Arifi:2024mff}, and the light-front holography~\cite{Brodsky:2003pw,Sharma:2023njj}.

In this work, we investigate exclusive $J/\psi$ production in photon-proton diffractive scattering using the Basis Light-Front Quantization (BLFQ) framework~\cite{Vary:2009gt,Vary:2025yqo}, a non-perturbative method for solving the relativistic many-body bound state problems within light-front quantum field theory~\cite{Zhao:2014xaa,Nair:2022evk,Wiecki:2014ola,li2016heavy,Lan:2019vui,Mondal:2019jdg,Xu:2021wwj,Lan:2021wok,Xu:2022yxb,Xu:2024sjt,Mondal:2025fdl,Qian:2020utg}. Recent advancements in BLFQ have enabled successful applications to various QCD observables, including electromagnetic form factors~\cite{Lan:2021wok,Mondal:2019jdg}, parton distribution functions~\cite{Lan:2021wok,Lan:2019img,Lan:2019rba,Lan:2019vui,Kaur:2024iwn,Mondal:2019jdg,Xu:2021wwj,Peng:2022lte}, and generalized parton distributions~\cite{Adhikari:2021jrh,Zhang:2023xfe,Lin:2023ezw,Liu:2022fvl,Xu:2021wwj}.

We calculate the exclusive $J/\psi$ production cross-section using the LFWFs of the proton and $J/\psi$ within the BLFQ framework, with photon LFWFs from perturbative QCD. The proton LFWFs are obtained from a light-front Hamiltonian in the valence Fock sector, incorporating a three-dimensional confinement potential and a one-gluon exchange interaction with fixed coupling~\cite{Mondal:2019jdg,Xu:2021wwj}. For the $J/\psi$, we first solve for the LFWFs from the light-front QCD Hamiltonian including both $|q\bar{q}\rangle$ and $|q\bar{q}g\rangle$ Fock components, together with three-dimensional confinement~\cite{Wu:2026gul}, then truncate the LFWFs to the $|q\bar{q}\rangle$ sector and renormalize them to unity. Following Ref.~\cite{Dumitru:2019qec}, the dipole scattering amplitude is evaluated using the proton wave function from BLFQ as the non-perturbative input. The cross-section is expressed as a convolution of the dipole amplitude with the photon and $J/\psi$ LFWFs. While our differential cross-section results are slightly lower than those in Ref.~\cite{Dumitru:2019qec}, the slopes are closely matched. These results provide initial conditions for the Balitsky-Kovchegov (BK) equation~\cite{Balitsky:1995ub,Kovchegov:1999yj} and support future EIC and EicC experiments.
\section{LFWFs of $J/\psi$ and proton in the BLFQ framework\label{Sec2}}
Following the Lepage-Brodsky convention~\cite{Brodsky1980}, the light-front variables are defined as $v^{\pm} \equiv v^0 \pm v^3$ and $\vec{v}_{\perp} \equiv (v_1, v_2)$. The LFWFs of bound states are obtained by solving the eigenvalue problem of the light-front Hamiltonian:  
$P^+ P^- |\Psi\rangle = M^2 |\Psi\rangle$,
where $P^+$ and $P^-$ denote the longitudinal momentum and the light-front Hamiltonian of the bound state, respectively. At fixed light-front time $x^+ \equiv x^0 + x^3$, the bound states with mass squared eigenvalue $M^2$ can be schematically expanded in terms of Fock sectors. Here, we illustrate this expansion for mesons and baryons as examples:
\begin{equation}
\begin{aligned}  &|\Psi\rangle_{\rm M}=\psi^{q\bar{q}}|q\bar{q}\rangle+\psi^{q\bar{q}g}|q\bar{q}g\rangle+\cdots,\\\
&|\Psi\rangle_{\rm B}=\psi^{qqq}|qqq\rangle+\psi^{qqqg}|qqqg\rangle+\cdots,\label{psi0}
\end{aligned}
\end{equation}
where $\psi^{\cdots}$ are the LFWFs corresponding to the Fock sectors $|\cdots\rangle$. In this work, the infinite Fock expansion is truncated for both the $J/\psi$ and the proton, retaining only their respective lowest Fock sector. Consequently, the $J/\psi$ is modeled as a quark-antiquark $|q\bar{q}\rangle$ state, and the proton as a three-quark $|qqq\rangle$ state.

For the proton, we adopt an effective light-front Hamiltonian $H_{\rm eff}=P_{\rm eff}^-P^+$ given by~\cite{Mondal:2019jdg,Xu:2021wwj},
\begin{equation}
\begin{aligned} \label{hami}
H_{\rm eff}=\sum_a \frac{{\vec k}_{\perp a}^2+m_{\rm q/KE}^2}{x_a}+\frac{1}{2}\sum_{a\ne b}V^{\rm conf}_{a,b}+\frac{1}{2}\sum_{a\ne b}V^{\rm OGE}_{a,b},
\end{aligned}
\end{equation}
where, $a$ and $b$ denote the index of particles in a Fock sector. $x_a$ and $\vec{k}_{\perp a}$ are the longitudinal fraction and transverse momentum of quark $a$, with $\sum_a x_a=1$ and $\sum_a \vec{k}_{\perp a}=0$. $V^{\rm conf}_{a,b}$ is the confining potential, which includes both the transverse and the longitudinal confinements. $V^{\rm OGE}_{a,b}$ represents the one-gluon exchange (OGE) interaction. They can be expressed as~\cite{Mondal:2019jdg,Xu:2021wwj}, 
\begin{equation}
\begin{aligned} \label{confOGE}
V^{\rm conf}_{a,b}&=\kappa_{\rm p}^4 x_ax_b \vec{r}^2_{\perp}+\frac{\kappa_{\rm p}^4}{4m^2_{\rm q/KE}}\partial_{x_a}(x_ax_b\partial_{x_b}), \\
V^{\rm OGE}_{a,b}&=\frac{F_{\rm c}~g_{s,{\rm p}}^2}{Q^2_{ab}} \bar{u}_{s'_a}(k'_a)\gamma^\mu{u}_{s_a}(k_a)\bar{u}_{s'_b}(k'_b)\gamma_\mu{u}_{s_b}(k_b).
\end{aligned}
\end{equation}
Here, $\kappa_{\rm p}$ is the confinement strength. The transverse separation between two quarks is $\vec{r}_{\perp} =\vec{r}_{\perp a} - \vec{r}_{\perp b}$, related to the holographic variable~\cite{brodsky2015light}. The OGE interaction involves the average momentum transfer squared $Q^2_{ab}=-q^2=-(1/2)(k'_a-k_a)^2-(1/2)(k'_b-k_b)^2$, color factor $F_{\rm c} = -2/3$ and coupling constant $g_{s,{\rm p}}$. ${u}_{s_a}(k_a)$ is the Dirac spinor of quark $a$ with spin $s_a$. To simulate effects of higher Fock sectors and additional QCD interactions, we use different quark masses in the kinetic energy ($m_{\rm q/KE}$) and OGE interaction ($m_{\rm q/OGE}$)~\cite{Mondal:2019jdg}.

For the $J/\psi$, we adopt an effective light-front Hamiltonian, $H_{\rm eff}=(P^-_{\rm{QCD}}+P^-_{\rm{C}})P^+$~\cite{Wu:2026gul,Lan:2021wok}, where $P^-_{\rm{QCD}}$ is the light-front QCD Hamiltonian, and $P^-_{\rm{C}}$ represents a model confining potential. 
In the light-front gauge $A^+=0$, the QCD Hamiltonian with one dynamical gluon is given by~\cite{Lan:2021wok,Brodsky:1997de},
\begin{align}
    P_{\rm{QCD}}^-&= \int \mathrm{d}x^- \mathrm{d}^2 x^{\perp} \Big\{\frac{1}{2}\bar{\psi}\gamma^+\frac{m_{0,c\bar{c}}^2+(i\partial^\perp)^2}{i\partial^+}\psi\nonumber\\
    &+\frac{1}{2}A_a^i\left[m_{g,c\bar{c}}^2+(i\partial^\perp)^2\right] A^i_a +g_{s,c\bar{c}}~\bar{\psi}\gamma_{\mu}T^aA_a^{\mu}\psi\nonumber\\
    &+ \frac{1}{2}g_{s,c\bar{c}}^2~\bar{\psi}\gamma^+T^a\psi\frac{1}{(i\partial^+)^2}\bar{\psi}\gamma^+T^a\psi \Big\}.
\end{align}
Here, $A^\mu_a$ and $\psi$ denote the gluon and quark fields, and $T^a = \lambda^a/2$ is half the Gell-Mann matrix. The charmonium system has coupling $g_{s,c\bar{c}}$, gluon mass $m_{g,c\bar{c}}$, and bare quark mass $m_{0,c\bar{c}}$. Although gluons are massless in QCD, we assign a phenomenological mass to fit low-energy spectra~\cite{Wu:2026gul}. A quark mass counterterm, $\delta m_{q,c\bar{c}} = m_{0,c\bar{c}} - m_{q,c\bar{c}}$, accounts for higher Fock sector corrections, with $m_{q,c\bar{c}}$ the renormalized mass. Following Refs.~\cite{Burkardt:1991tj,Glazek:1992aq}, we introduce an independent vertex quark mass $m_{f,c\bar{c}}$, tuned to reproduce the $\eta_c$–$J/\psi$ mass splitting, the $J/\psi$ decay constant and form factor, and consistency with related experimental data~\cite{Wu:2026gul}.

The confining potential consists of the transverse and longitudinal terms. 
In the valence Fock sector, it is written as~\cite{Lan:2021wok,li2016heavy},
\begin{equation}
  \begin{aligned}\label{eqn:Hc}
  &P_{\rm{C}}^-P^+=\kappa_{c\bar{c}}^4\left\{x(1-x) \vec{r}_{\perp}^2-\frac{\partial_{x}[x(1-x)\partial_{x}]}{(m_{q,c\bar{c}}+m_{\bar{q},c\bar{c}})^2}\right\},
  \end{aligned}
\end{equation}
where $\kappa_{c\bar{c}}$ is the strength coefficient of confinement in the charmonium system, and again $\vec r_{\perp}=(\vec r_{\perp q}-\vec r_{\perp \bar{q}})$ stands for the transverse separation between two quarks. 
We exclude an explicit confinement interaction in the $|qqqg\rangle$ sector. Instead, confinement effects are effectively modeled through the transverse basis truncation and the introduction of an effective gluon mass, in analogy with functional approaches where gluons dynamically acquire effective masses~\cite{Cornwall:1981zr,Alkofer:2000wg,Deur:2016tte}.

 Using the BLFQ framework~\cite{Vary:2009gt}, we solve the Hamiltonian equation in a chosen basis space. The Fock sectors in Eq.~(\ref{psi0}) are taken to be direct products of single particle states $|\alpha\rangle=\otimes_i|\alpha_i\rangle$. We employ the discretized light-cone quantization  basis~\cite{Brodsky:1997de} and a two-dimensional harmonic oscillator (2D-HO) basis functions to describe the longitudinal and transverse dynamics of single-particle states, respectively. More specifically, in the longitudinal direction, the single particle is confined in a one-dimensional box of length $2L$ with periodic (anti-periodic) boundary conditions for the boson (fermion). For the $i\rm{th}$ single-particle state, the longitudinal momentum is discretized as $p_i^+=\frac{2\pi}{L}k_i$, where the longitudinal quantum number $k_i$ is a integer (half-integer) for bosons (fermions). Here, we neglect the zero mode $k_i=0$ for the boson. In the transverse plane, the 2D-HO wave function $\Phi_{n_im_i}(\vec{p}_{i\perp}, b)$ carries the radial and the angular quantum numbers denoted by $n_i$ and $m_i$, respectively. $\vec{p}_{i\perp}$ is the transverse momentum of the $i\rm{th}$ particle. $b$ is the HO basis scale parameter. Each single-particle state $|\alpha_i\rangle$ is characterized by four quantum numbers, $|\alpha_i\rangle=|k_i,n_i,m_i,\lambda_i\rangle$, where $\lambda_i$ is the light-front helicity. In this work, each Fock sector ($|q\bar{q}\rangle$ and $|q\bar{q}g\rangle$ for $J/\psi$, and $|qqq\rangle$ for proton) allows for a unique color-singlet state.

The projection of total angular momentum $M_J$ consisting of orbital angular momentum projection plus spin projection is conserved on the light front. Meanwhile, we respectively introduce truncation parameters $N_{\rm{max}}$ and $K$ in transverse and longitudinal directions to perform the numerical calculation.
These quantities satisfy the following conditions, 
\begin{equation}
\begin{aligned}
    \begin{cases}
          M_J=\sum_i (m_i+\lambda_i),\\
          N_{\rm{max}}\ge \sum_i(2n_i+|m_i|+1),\\
          K=\sum_{i}k_{i}.
    \end{cases}
\end{aligned}
\end{equation}
For the $i\rm{th}$ particle, the longitudinal momentum fraction is defined as $x_i=p_i^+/P^+=k_i/K$. Therefore, $K$ serves as the longitudinal truncation parameter and determines the resolution of the longitudinal distributions. The $N_{\rm{max}}$ truncation introduces the ultraviolet (UV) and infrared (IR) cutoffs of the transverse basis. In momentum space, the UV cutoff $\Lambda_{\rm{UV}} \simeq b\sqrt{N_{\rm{max}}}$ and the IR cutoff $\Lambda_{\rm{IR}} \simeq b/\sqrt{N_{\rm{max}}}$ ~\cite{Zhao:2014xaa}. 

Through diagonalizing the light-front Hamiltonian, we obtain the mass spectra $M^2$ and the corresponding eigenvectors $\psi_{M_J}^{\mathcal{N}}(\{\alpha_i\})$, which can be converted to the LFWFs in momentum representation, 
\begin{equation}
\begin{aligned}
  &\psi_{M_J,\{\lambda_i\}}^{\mathcal{N}}({\{x_i,\vec{p}_{i\perp}\}})\\
  &=\sum_{ \{n_i m_i\} }\psi_{M_J}^{\mathcal{N}}(\{\alpha_i\})\prod_{i=1}^{\mathcal{N}}  \Phi_{n_i m_i}(\vec{p}_{i\perp},b)\,,
\label{eqn:wf}
\end{aligned}
\end{equation}
where $\mathcal{N}$ is the total number of particles in the Fock sector. $\alpha_i$ represents the four quantum numbers $k_i$, $n_i$, $m_i$ and $\lambda_i$, characterizing the $i\rm{th}$ particle as mentioned before.

For the $J/\psi$, with the truncation $\{N_{\rm{max}}, K\}=\{12,17\}$, we determine the parameters summarized in Table~\ref{para2} by fitting the masses of low-lying charmonium systems spectra~\cite{Wu:2026gul}. For the proton, with the truncation $\{N_{\rm{max}}, K\}=\{10,16.5\}$, we determine the parameters summarized in Table~\ref{para1} by fitting the proton mass and its electromagnetic properties~\cite{Mondal:2019jdg,Xu:2021wwj}.
The resulting proton LFWFs have been successfully used to study a range of properties, including electromagnetic and axial form factors, radii, PDFs, GPDs, TMDs, GFFs, and angular momentum distributions~\cite{Mondal:2019jdg,Xu:2021wwj,Liu:2022fvl,Hu:2022ctr,Kaur:2023lun,Liu:2024umn,Nair:2024fit}, while the $J/\psi$ LFWFs have simultaneously described its PDFs, decay constant, charge radius, radiative transitions, and electromagnetic form factors~\cite{Wu:2026gul,Maris:2020wew,Li:2018uif}.

\begin{table}[ht]
  \caption{Model parameters for the basis truncations $\{N_{\text{max}},K\}=\{12,17\}$ for the $J/\psi$~\cite{Wu:2026gul}. All are in units of GeV except $g_{s,c\bar{c}}$.}
  \vspace{0.15cm}
  \label{para2}
  \centering
    \setlength{\tabcolsep}{1mm}{
  \begin{tabular}{cccccc}
    \hline\hline
         $m_{\rm q,c\bar{c}}$ & $m_{\rm g,c\bar{c}}$ &$b_{\rm c\bar{c}}$ &$\kappa_{\rm c\bar{c}}$ &$m_{\rm f,c\bar{c}}$  &$g_{\rm s,c\bar{c}}$ \\        
    \hline 
        1.54 & 0.50 & 1.23 &1.23 & 5.04 & 2.24 \\       
    \hline\hline
  \end{tabular}}
\end{table}

\begin{table}[ht]
  \caption{Model parameters for the basis truncations $\{N_{\text{max}},K\}=\{10,16.5\}$ for the proton~\cite{Mondal:2019jdg,Xu:2021wwj}. All are in units of
GeV except $g_{s,{\rm p}}$.}
  \vspace{0.15cm}
  \label{para1}
  \centering
    \setlength{\tabcolsep}{1mm}{
  \begin{tabular}{ccccccc}
    \hline\hline
         $m_{\rm q/KE}$ & $m_{\rm q/OGE}$ & $b_{\rm p}$ &$\kappa_{\rm p}$  &$g_{\rm s, p}$ \\        
    \hline 
        0.30 & 0.20 & 0.34 &0.34 &3.72 \\       
    \hline\hline
  \end{tabular}}
\end{table}
\section{Differential cross section for exclusive $J/\psi$ production\label{Sec3}}

In the high-energy limit, the differential cross section for exclusive $J/\psi$ production is expressed as~\cite{kowalski2006exclusive},
\begin{equation} \label{eq:sigma_AA*}
\frac{{\rm d}\sigma}{{\rm d} t} = \frac{1}{16\pi} \sum_{T,L}
\left| {\cal A}_{T,L}^{\gamma^* p \rightarrow M\, p} \right|^2 \, ,
\end{equation}
where $t$ is the total momentum transfer squared. On the right-hand side, the squared amplitude is independent of the direction of $\vec K_\perp$ due to rotational invariance, and can thus be evaluated for any chosen orientation.

The amplitude for this process is given by~\cite{kowalski2006exclusive},
\begin{equation}
\begin{aligned} \label{eq:ATL_JPsi_P}
&{\cal A}_{T,L}^{\gamma^*p\rightarrow J/\Psi~p} (Q^2,\vec K_\perp) \\ &= i \int
{\rm d}^2 \vec r_\perp \int\limits_0^1 \frac{dz}{4\pi} \, \left(\Psi_{\gamma^*}
\Psi^*_{J/\Psi}\right)_{T,L}(\vec r_\perp,z,Q^2)\\ & \times e^{-i\frac{(1-2z)}{2}\vec r_\perp\cdot \vec K_\perp} \,\, {\cal T}(\vec r_\perp,\vec K_\perp)\,,
\end{aligned}
\end{equation}
where $\Psi_{J/\psi}$ and $\Psi_{\gamma^*}$ denote the $J/\psi$ and virtual photon LFWFs, with $T$ and $L$ referring to transverse and longitudinal polarizations, respectively. Their product is summed over the helicities of the $c$ and $\bar{c}$ quarks. $z$ is the longitudinal momentum fraction of the $c$ quark. For $\Psi_{J/\psi}$, we use the expressions in Eq.~\eqref{eqn:wf} for numerical calculations, while $\Psi_{\gamma^*}$ is obtained from perturbative QCD. 
For a longitudinally polarized virtual photon, the wave function reads~\cite{kowalski2006exclusive},
\begin{equation}
\begin{aligned}
  \Psi^{\gamma^*}_{h\bar{h},\lambda=0}(\vec r_\perp,z,Q^2) = 2 e_f e \, 
  \delta_{h,-\bar h} \, Qz(1-z)\, \frac{K_0(\epsilon |\vec r_\perp|)}{2\pi},
  \label{long_photon_wf}
\end{aligned}
\end{equation}
where $e_f$ denotes the fractional electric charge of quark and $e=\sqrt{4\pi\alpha}$ is the elementary charge, with $\alpha$ the fine-structure constant.
While for transverse polarization, it is given by
\begin{equation}
\begin{aligned}
  &\Psi^{\gamma^*}_{h\bar{h},\lambda=\pm}(\vec r_\perp,z,Q^2) =
  \sqrt{2}\lambda\, e_f e \,
  \Big\{
  ie^{\lambda\, i\phi_r}\big[
    z\delta_{h,\lambda}\delta_{\bar h,-\lambda} \\ &- 
    (1-z)\delta_{h,-\lambda}\delta_{\bar h,\lambda}\big] \partial_r + \, 
  m_q \delta_{h,\lambda}\delta_{\bar h,\lambda}
  \Big\}\, \frac{K_0(\epsilon |\vec r_\perp|)}{2\pi}.
  \label{transv_photon_wf}
\end{aligned}
\end{equation}
Here, $m_q$ is the quark mass, $K_0$ is the modified Bessel function of the second kind, $\phi_r$ is the azimuthal angle of $\vec r_\perp$, and $\epsilon^2 \equiv z(1-z)Q^2 + m_q^2$.

\begin{figure*}[htp]
 \centering
   \includegraphics[width=6in,height=1.5in]{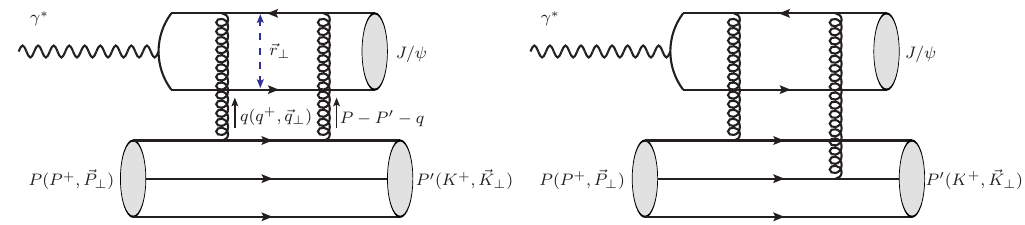}  
   \caption{Leading-order Feynman diagrams contributing to $J/\psi$ production via ${\cal C}$-even two-gluon exchange. The left and right diagrams illustrate, respectively, the cases that two color charges act on the same quark in the proton and on two different quarks in the proton.}
   \label{Feynman_diagram}
\end{figure*}
\begin{figure*}[htp]
  \centering
    \includegraphics[width=0.45\textwidth]{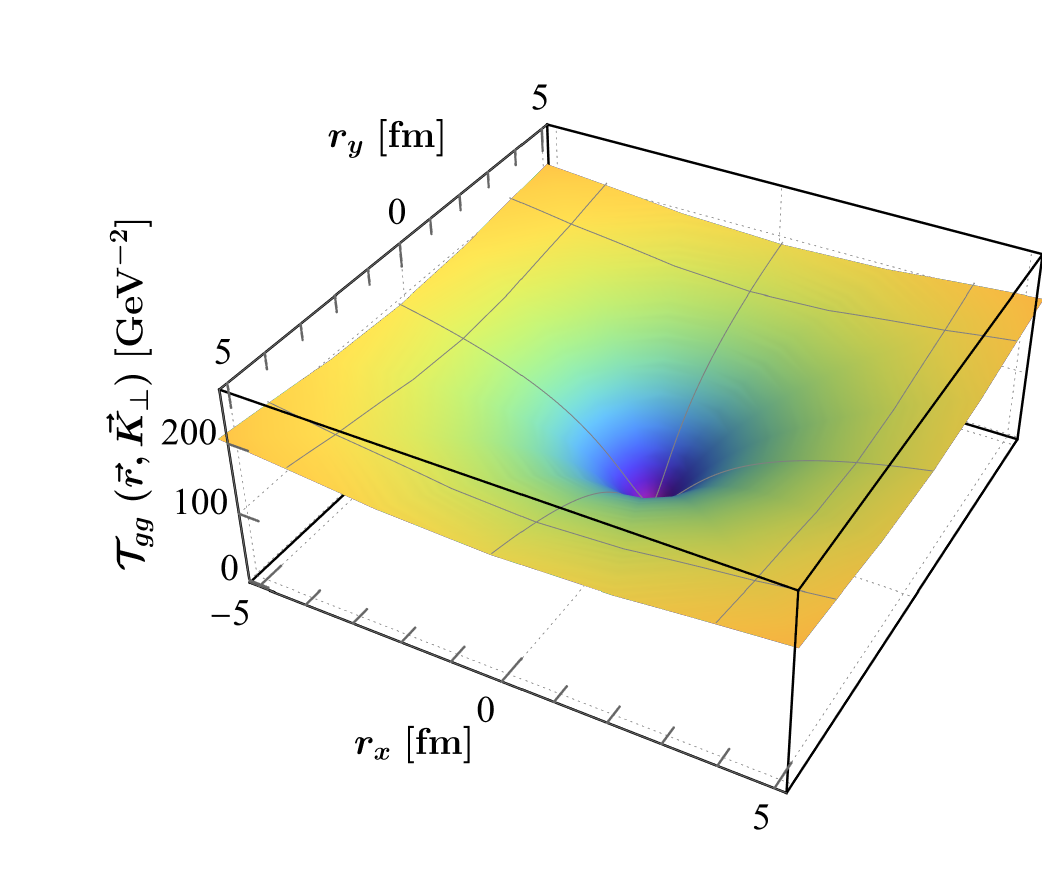}
    \includegraphics[width=0.45\textwidth]{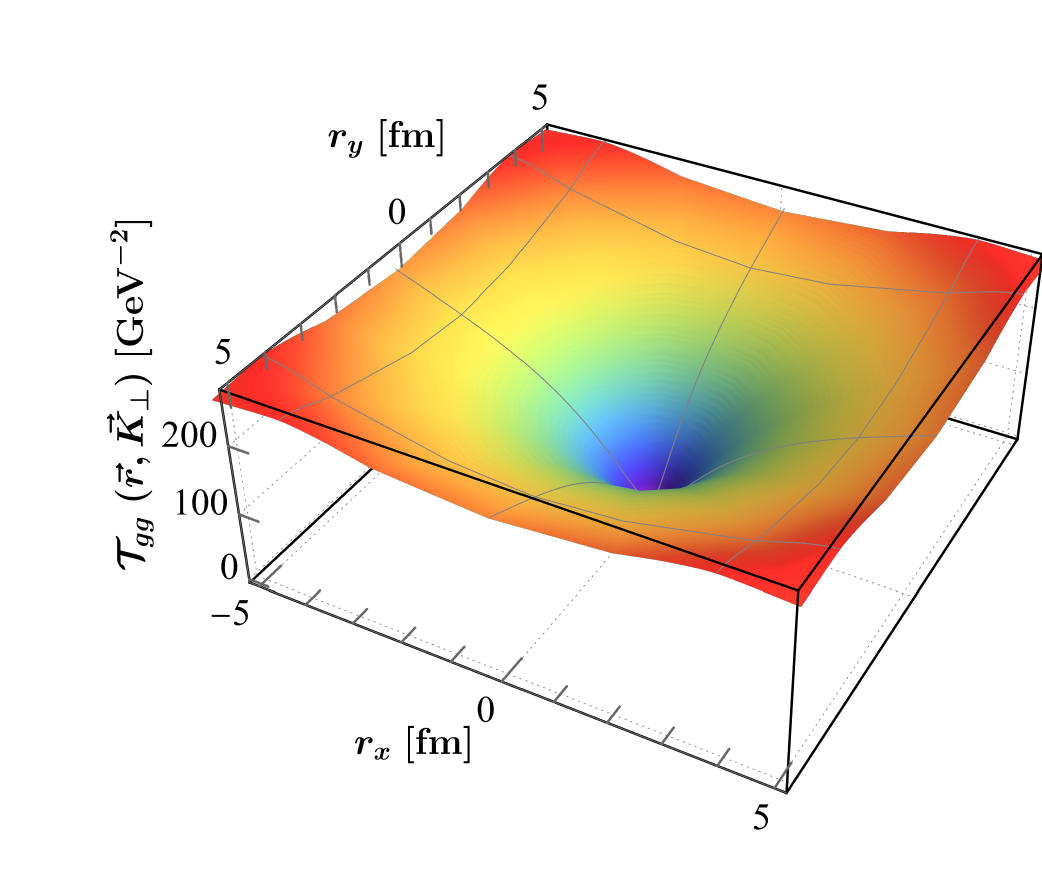}\\
        \includegraphics[width=0.45\textwidth]{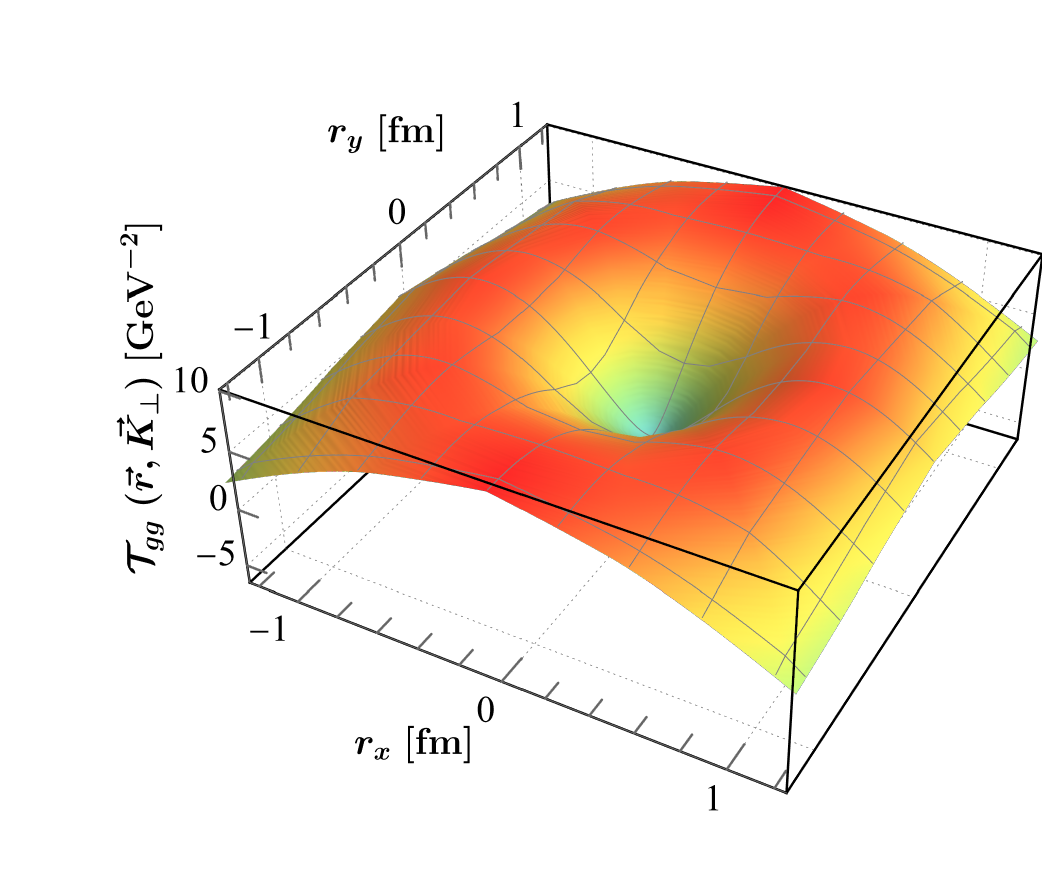}
    \includegraphics[width=0.45\textwidth]{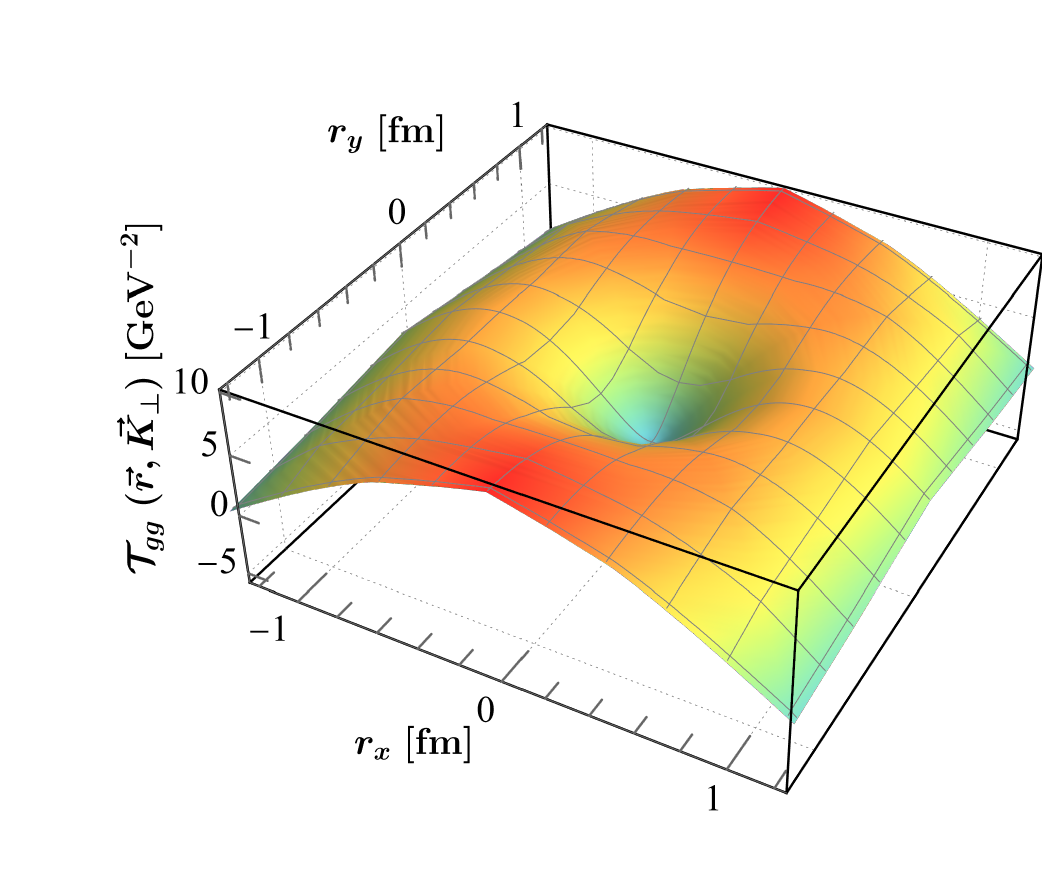}
\caption{{The dipole scattering amplitude ${\cal T}_{gg}$ at $\vec{K}^2_{\perp}=0$ (upper panels) $\vec{K}^2_{\perp}=0.5~\rm{GeV}^2$ (lower panels), computed with two different LFWFs of the proton. The left panels show the results obtained from the "harmonic oscillator" proton wave function~\cite{brodsky1994wavefunction}. The right panels show the results obtained from the BLFQ proton wave function~\cite{Mondal:2019jdg}.}}
      \label{comparsion_tgg_3D}
\end{figure*}
\begin{figure*}[htp]
  \centering
      \includegraphics[width=0.45\textwidth]{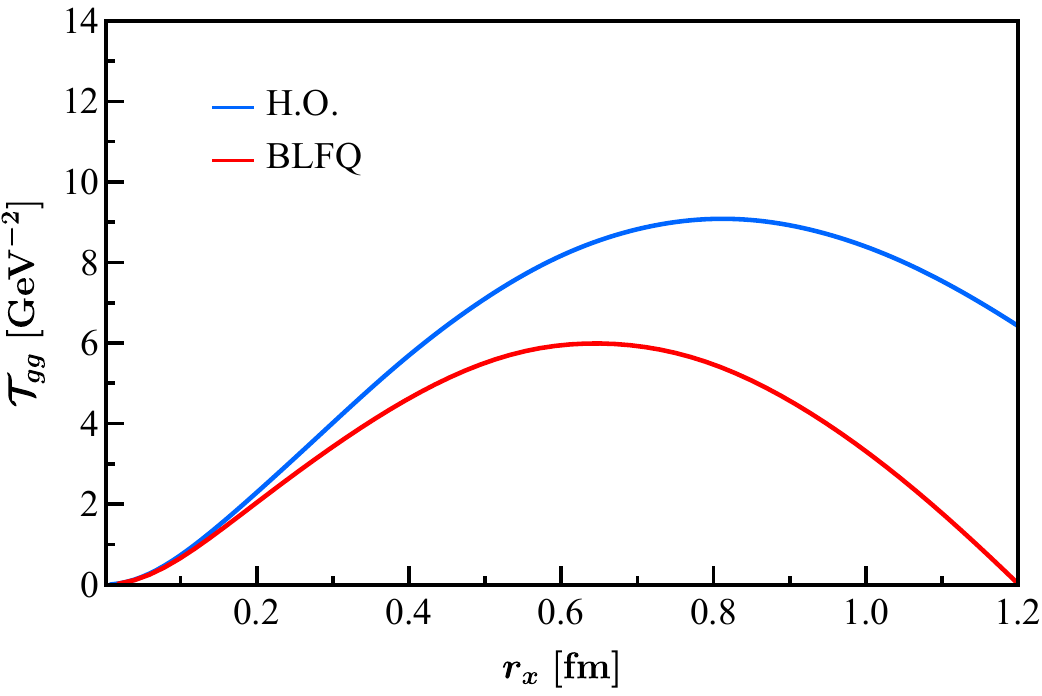}
      \includegraphics[width=0.45\textwidth]{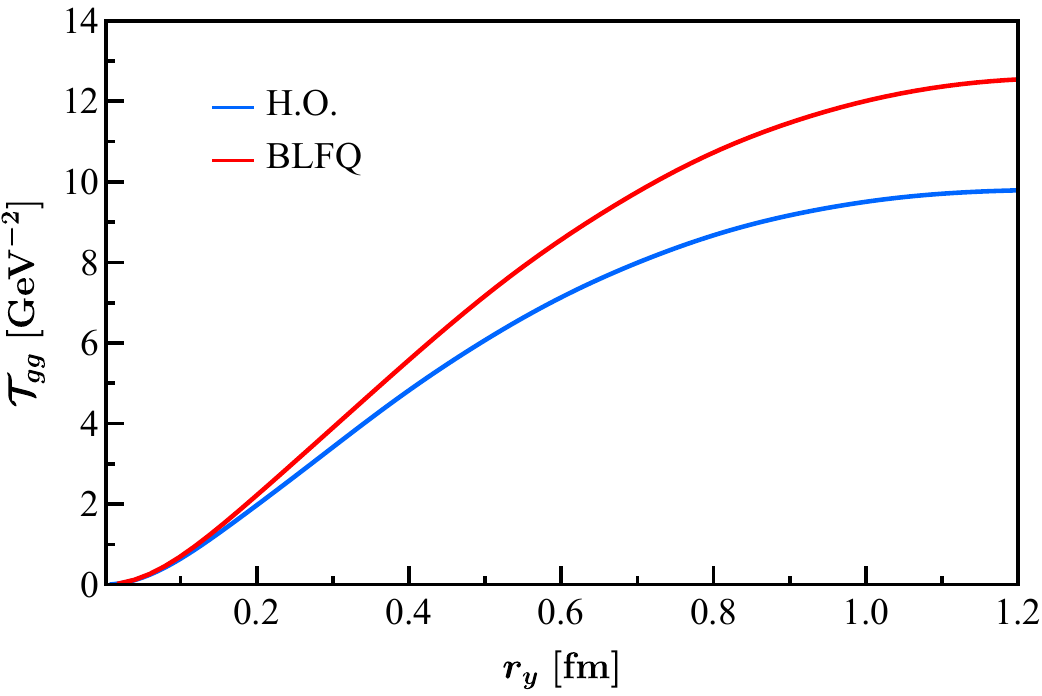}
\caption{Comparison of the dipole scattering amplitude ${\cal T}_{gg}$ at $\vec{K}_{\perp}^2 = 0.5~\rm{GeV}^2$, obtained using the “harmonic oscillator (H.O.)”~\cite{brodsky1994wavefunction} and BLFQ proton LFWFs~\cite{Mondal:2019jdg}. Left panel: Variation of ${\cal T}_{gg}$ along the $r_x$ direction ($\vec{K}_{\perp}$ direction). Right panel: Variation along the $r_y$ direction.}
      \label{comparsion_tgg}
\end{figure*}
\begin{figure*}[htp]
  \centering
    \includegraphics[width=0.45\textwidth]{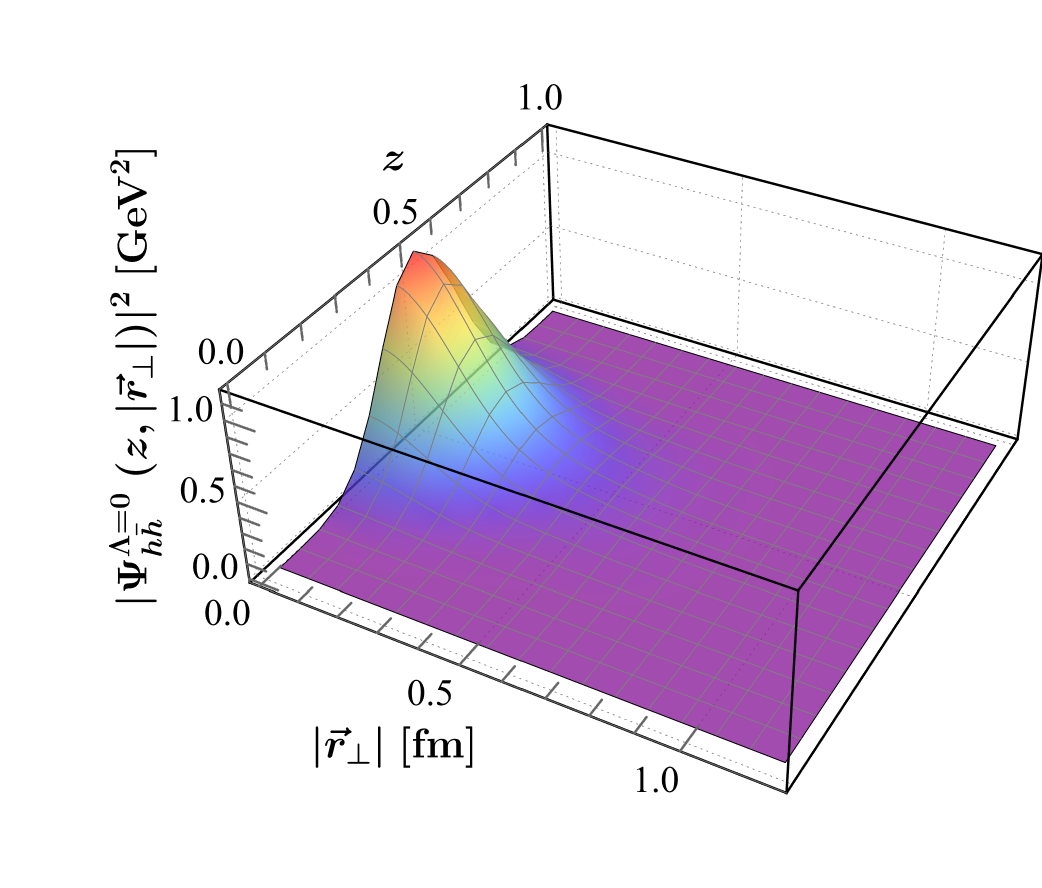}
    \includegraphics[width=0.45\textwidth]{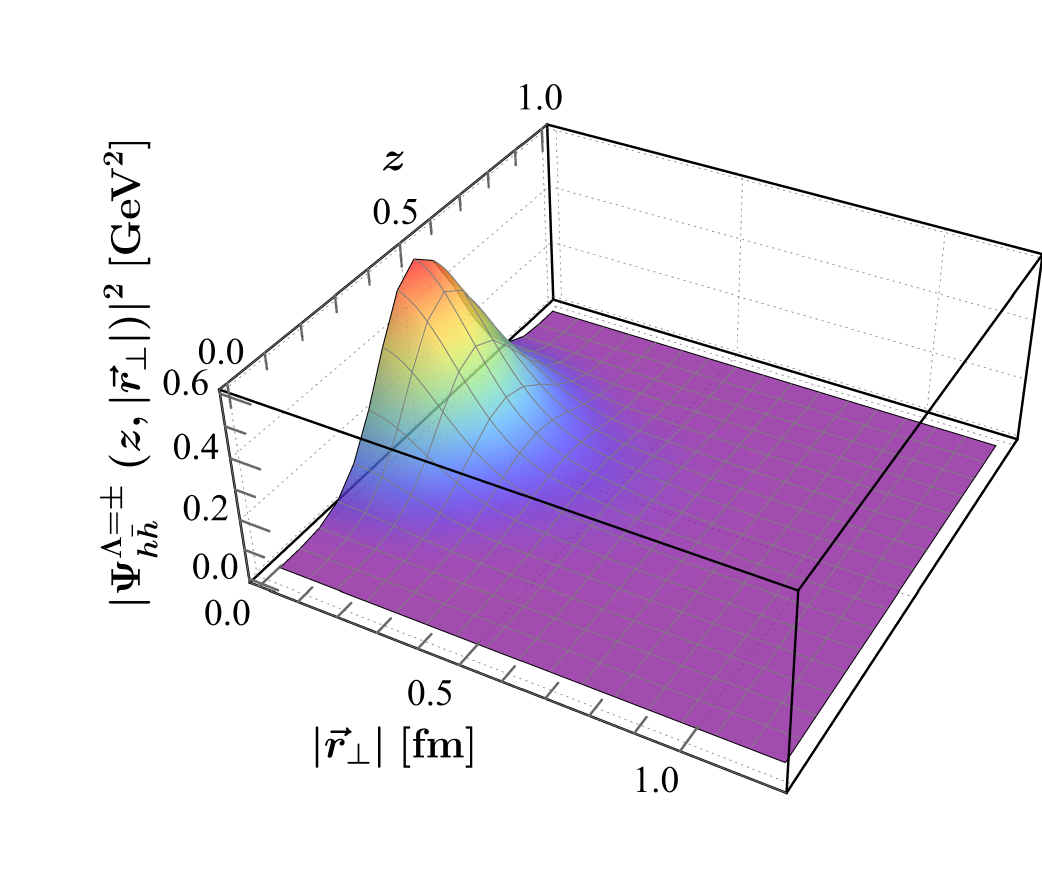}
\caption{Three-dimensional distributions of the BLFQ $J/\psi$ LFWFs for longitudinally (left) and transversely (right) polarized states, shown as functions of the quark longitudinal momentum fraction $z$ and the dipole separation $|\vec{r}_\perp|$ (in fm).}
      \label{jpsiwf_3D}
\end{figure*}

\begin{figure*}[htp]
  \centering
      \includegraphics[width=0.432\textwidth]{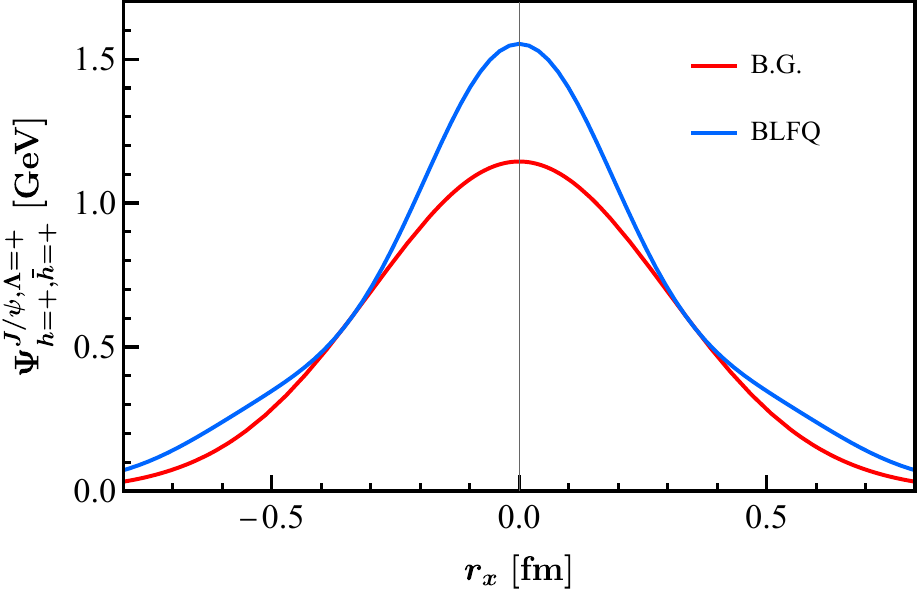} 
      \qquad
      \includegraphics[width=0.45\textwidth]{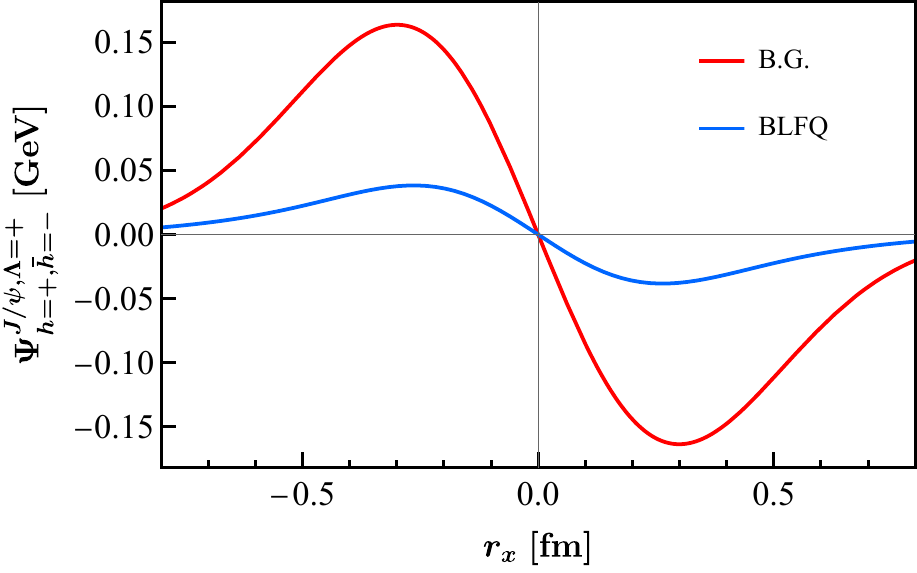}
      \caption{The S- and P-wave $J/\psi$ LFWFs $\Psi^{J/\psi,\Lambda=+}_{h=+,\bar{h}=+}$ and $\Psi^{J/\psi,\Lambda=+}_{h=+,\bar{h}=-}$, computed using the “Boosted Gaussian (B.G.)” model and the BLFQ framework, as  functions of the transverse dipole separation $\vec r_\perp$ taken along the $x$-axis ($r_y=0$). Left panel:  $\Psi^{J/\psi,\Lambda=+}_{h=+,\bar{h}=+}$. Right panel: $\Psi^{J/\psi,\Lambda=+}_{h=+,\bar{h}=-}$. The blue and red lines correspond to the BLFQ and B.G. model results  for $z=0.5$, respectively.}
      \label{comparsion_jpsiwf}
\end{figure*}

The invariant amplitude ${\cal T}$ for the elastic scattering of a $c\bar{c}$ pair off the target proton is given by~\cite{Dominguez:2011wm},
\begin{align}
   {\cal T} (\vec r_\perp, \vec K_\perp) =& \int {\rm d}^2 \vec b_\perp \, e^{i \vec b_\perp \cdot
     \vec K_\perp} \, {\cal T} (\vec r_\perp,\vec b_\perp; \vec K_\perp)~, \label{eq:Tij} 
     \end{align}
where
\begin{align}
   &{\cal T} (\vec r_\perp,\vec b_\perp; \vec K_\perp)\nonumber \\
   & = 2\, N_c\, \bigg[ 1 - \frac{1}{N_c}\,{\rm tr} \,\bigg< U\bigg(\vec b_\perp +
     \frac{\vec r_\perp}{2}\bigg) 
      \, U^\dagger\bigg( \vec b_\perp - \frac{\vec r_\perp}{2}\bigg)\bigg>_{\vec
       K_\perp} \bigg]\, . \label{eq:Tamplitude_dipole_Kt}
\end{align}
Here, the lightlike Wilson line is defined as
$U^\dagger(\vec{x}_T)=\mathcal{P}\exp\!\left[ig\int dx^- A^+(x^-,\vec{x}_T)\right]$,
with path ordering in the $x^-$ direction. For exclusive $J/\psi$ production, the leading-order contribution arises from ${\cal C}$-even two-gluon (Pomeron) exchange, as illustrated in Fig.~\ref{Feynman_diagram}. This corresponds to expanding Eq.~(\ref{eq:Tamplitude_dipole_Kt}) to order $(gA^+)^2$. Using the Poisson equation $\Delta_\perp^2 A^{+a}=-g\rho^a$, we then obtain the invariant amplitude $\mathcal{T}$ for this process as~\cite{Dumitru:2019qec,dumitru2018extracting},
\begin{eqnarray}
{\cal T}_{gg}(\vec r_\perp,\vec K_\perp) &=& 
  -\,\int\limits_q
  \frac{(4\pi\alpha_s)^2 N_c C_F}{(\vec q_\perp -\frac{1}{2}\vec K_\perp)^2 \, (\vec q_\perp + \frac{1}{2} \vec
  K_\perp)^2} \nonumber\\  
  & & \times \, \left(\cos\left(\vec r_\perp \cdot {\vec q_\perp}\right) 
- \cos\left(\frac{{\vec r_\perp}\cdot \vec K_\perp}{2}\right)\right) \nonumber\\
& & \times \; \; G\left({\vec q_\perp}-\frac{1}{2}\vec K_\perp,-{\vec q_\perp}-\frac{1}{2}\vec K_\perp\right),    \label{eq:Pomeron}
\end{eqnarray}
with the shorthand notation $\int_q = \int{\rm d}^2 \vec q_\perp/(2\pi)^2$. $N_c = 3$ is the dimension of the SU(3) color fundamental representation and $C_F = 4/3$. The strong coupling is $\alpha_s = g_s^2/(4\pi)$. $\rho^a \equiv \bar{\psi}_i\gamma^+\psi_j(t^a)_{ij}$ is the color charge density operator.

$G$ can be extracted from the correlator of two color charge density operators~\cite{Dumitru:2019qec,dumitru2018extracting},
\begin{eqnarray}
& &\langle \, \rho^a(\vec q_\perp) \, \rho^b(-\vec q_\perp-\vec K_\perp) \,\rangle_{\vec K_\perp} \nonumber\\
&=&\,\frac{1}{2}\,\delta^{ab}\,\int {\rm d} x_1 {\rm d}  x_2
{\rm d} x_3 \, \delta(1-x_1-x_2-x_3) \nonumber\\
& &\times  \int \frac{{\rm d}^2 \vec{p}_{1\perp} {\rm d}^2 \vec{p}_{2\perp} {\rm d}^2 \vec{p}_{3\perp}}{(16\pi^3)^2}
\, \delta(\vec{p}_{1\perp}+\vec{p}_{2\perp}+\vec{p}_{3\perp}) \nonumber\\
& &\psi_3(\vec p_{1\perp}, \vec p_{2\perp}, \vec p_{3\perp}) \left[\psi_3^*(\vec p_{1\perp} + (1-x_1) \vec K_\perp,\right. \nonumber\\
& &\vec p_{2\perp} -x_2 \vec K_\perp, \vec p_{3\perp} -x_3 \vec K_\perp) - \psi_3^*(\vec p_{1\perp} +\vec q_\perp + \nonumber\\
& &\left. (1-x_1)\vec K_\perp, \vec p_{2\perp} -\vec q_\perp -x_2 \vec K_\perp, \vec p_{3\perp} -x_3 \vec K_\perp) \right] \nonumber\\
&\equiv& \frac{1}{2}\delta^{ab}\, G(\vec q_\perp, -\vec q_\perp - \vec K_\perp)~.
\label{eq:rho2_Kt}
\end{eqnarray}
Here, $\langle \cdots \rangle_{\vec{K}_{\perp}} \equiv \langle K| \cdots|P\rangle/\langle K|P\rangle$ denotes the expectation value of the operators between the initial proton state with momentum $P$ and the final proton state with momentum $K$. The transverse momentum of the initial proton state $\vec{P}_{\perp}=\sum^3_{i=1} \vec{p}_{i\perp}=0$, $\vec{K}_{\perp}$ is the transverse momentum of the final proton state, $\vec q_\perp$ is the transverse momentum carried by one of the gluons. $G(\vec q_\perp, -\vec q_\perp - \vec K_\perp)$ is invariant under rotations that preserve the relative angle between $\vec q_\perp$ and $\vec K_\perp$, and under exchange of its arguments, $G(\vec q_\perp, -\vec q_\perp - \vec K_\perp) = G(-\vec q_\perp - \vec K_\perp, \vec q_\perp)$. It vanishes when either $\vec q_\perp$ or $-\vec q_\perp - \vec K_\perp$ approaches zero, reflecting the proton’s color charge neutrality. $\psi_3$ in Eq.~(\ref{eq:rho2_Kt}) denotes the proton LFWF in the valence Fock sector $|qqq\rangle$. While Ref.~\cite{brodsky1994wavefunction} employed ``harmonic oscillator" and ``power-law" model wave functions for numerical estimates, we use the proton LFWFs obtained within the BLFQ framework.
\begin{figure*}[ht]
  \centering
      \includegraphics[width=4in,height=2.6in]{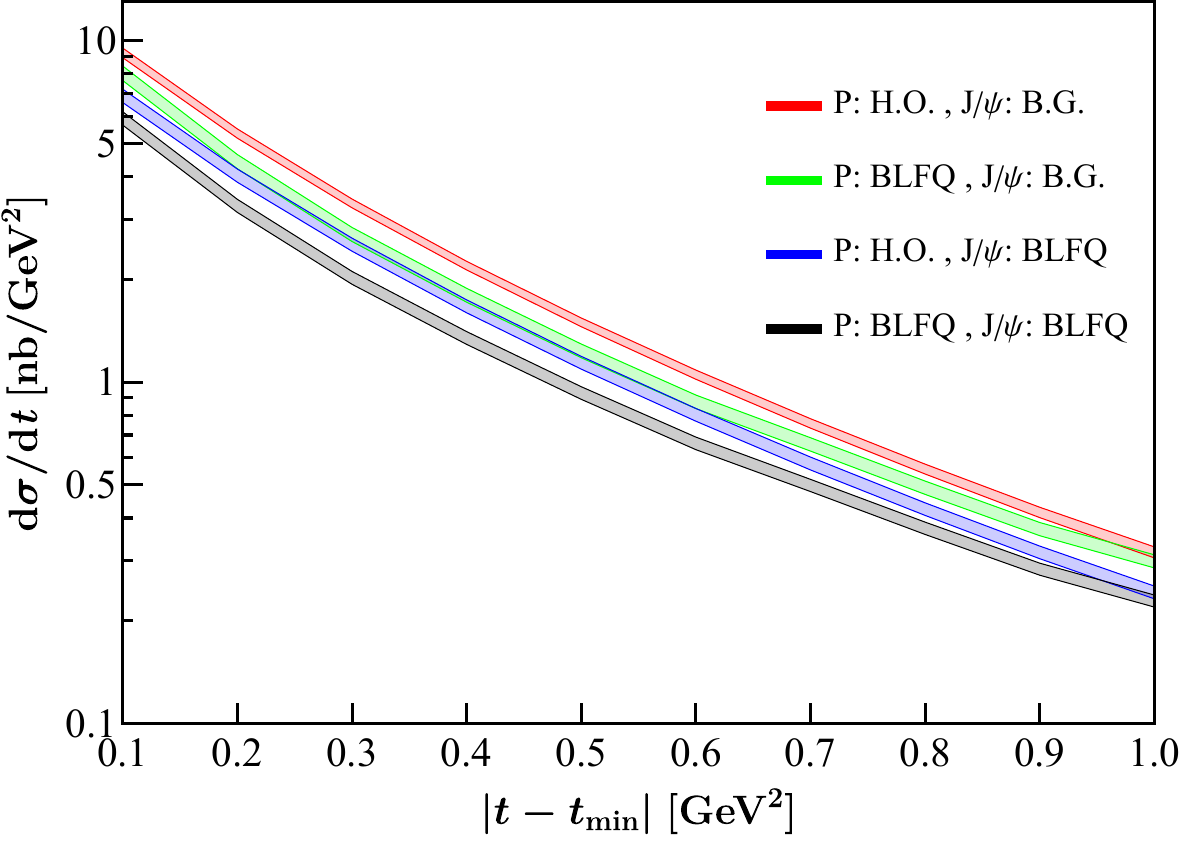}
      \caption{The differential cross section for exclusive $J/\psi$ production. The bands indicate the variation covering the range $0<Q^2 < 0.5~\rm{GeV^2}$. The red line represents the results calculated entirely using the wave function employed by Dumitru et al.~\cite{Dumitru:2019qec}. The green line corresponds to the results obtained by replacing the proton wave function with the BLFQ wave function. The blue line indicates the results calculated by replacing the $J/\psi$ wave function with the BLFQ wave function. The black line shows the results obtained when both the proton and $J/\psi$ wave functions are replaced with the BLFQ wave function.}
      \label{Fig_1_of_tff}
\end{figure*}
\section{Numerical results\label{Sec4}}
The exclusive $J/\psi$ production cross section in photon-proton diffractive scattering is computed using three wave functions: the photon wave function, obtained analytically via perturbative QCD, and the hadronic wave functions of the proton and $J/\psi$, computed numerically using the BLFQ framework.



Before presenting the differential cross section for $J/\psi$ photoproduction, we discuss the dipole scattering amplitude ${\cal T}_{gg}$, a key ingredient that depends on the proton LFWFs. 
Figure~\ref{comparsion_tgg_3D} shows the dipole scattering amplitude ${\cal T}_{gg}$ as a function of the transverse dipole vector $\vec{r}_\perp = (r_x, r_y)$. For zero transverse momentum transfer ($|\vec{K}_\perp| = 0$), illustrated in the upper panels of Fig.~\ref{comparsion_tgg_3D}, the amplitude is rotationally symmetric, with a central minimum at $\vec{r} = 0$ surrounded by a ring of maxima, reflecting the color neutrality of the proton. For finite transverse momentum transfer ($|\vec{K}_\perp| = \sqrt{0.5}~\rm{GeV}$ oriented along the $\hat{x}$-direction), shown in lower panels of Fig.~\ref{comparsion_tgg_3D}, the amplitude becomes anisotropic: the maxima are elongated along the $r_y$ direction, while the amplitude falls off more rapidly along the $r_x$ direction.

Figure~\ref{comparsion_tgg} compares our results with the dipole amplitude computed using the ``harmonic oscillator" wave function from Ref.~\cite{Dumitru:2019qec}. The BLFQ amplitude exhibits greater sensitivity to $\vec{K}_\perp$, with a sharper fall-off along $r_x$, while the behavior along $r_y$ remains similar. This pattern of a central minimum surrounded by maxima highlights the spatial structure of the proton in the valence Fock sector and directly influences the exclusive $J/\psi$ production cross section.

Another key ingredient is the $J/\psi$ wave function. Figure~\ref{jpsiwf_3D} shows the three-dimensional probability distributions of the $J/\psi$ LFWFs, $|\Psi_{h,\bar{h}}^\Lambda (z, \vec{r}_{\perp})|^2$, as functions of the longitudinal momentum fraction $z$ and transverse dipole separation $\vec{r}_{\perp}$, for both longitudinally and transversely polarized $J/\psi$. The BLFQ LFWFs in the leading Fock component peak at $z = 0.5$ and $\vec{r}_{\perp} = 0$, and smoothly decrease as $z \to 0,1$ and as $\vec{r}_{\perp}$ increases. 


We further compare our $J/\psi$ LFWFs in the BLFQ framework with the “Boosted Gaussian” model~\cite{kowalski2006exclusive} in Fig.~\ref{comparsion_jpsiwf}. We observe that the S-wave component of the BLFQ wave functions dominates over the phenomenological model, while the P-wave component is smaller in magnitude compared to the “Boosted Gaussian” wave functions.

Figure~\ref{Fig_1_of_tff} shows our results for the differential cross section of $J/\psi$ production as a function of $|t-t_{\rm min}|$. Here, $t$ denotes the total squared momentum transfer, while $t_{\rm min}$ is the minimum kinematically allowed value arising from the longitudinal momentum transfer in the $\gamma^*+p\rightarrow J/\psi+p$ process. A nonzero $t_{\rm min}$ is required because the proton must provide a longitudinal momentum transfer to convert a virtual photon with finite $Q^2$ into a heavy charmonium state. In our calculation, we employ the eikonal approximation and neglect the explicit longitudinal momentum transfer. Since the transverse momentum transfer $-\vec{K}_{\perp}^{\,2}$ dominates in this regime, we approximate the longitudinal contribution by $t_{\rm min}$, such that $t-t_{\rm min}$ can be identified with $-\vec{K}_{\perp}^{\,2}$.
The red band shows results calculated entirely with the LFWFs used by Dumitru and Stebel~\cite{Dumitru:2019qec}, while the green and blue bands correspond to replacing the proton and the $J/\psi$ LFWFs with their BLFQ counterparts, respectively. The black band shows the result when both the proton and $J/\psi$ LFWFs are replaced by BLFQ wave functions.
Fitting the cross sections with an exponential falloff ${\rm d}\sigma/{\rm d}t \propto e^{B(t-t_{\rm min})}$ in the range $-1~\rm{GeV^2} < t-t_{\rm min}  < -0.5~\rm{GeV^2}$ gives slopes $B \approx 3~\rm{GeV^{-2}}$ for all four cases, consistent with experimental data at comparable energies~\cite{camerini1975photoproduction}. 


The origin of the differences between our results and those reported by Dumitru and Stebel~\cite{Dumitru:2019qec} can be traced to the respective ${\cal T}_{gg}$ and $J/\psi$ LFWFs, as illustrated in Figs.~\ref{comparsion_tgg} and~\ref{comparsion_jpsiwf}. 
The BLFQ results for ${\cal T}_{gg}$ are more sensitive to $\vec{K}_\perp$ and exhibit a faster fall-off along the $r_x$ ($\vec{K}_\perp$) direction. 
For $0\le Q^2 \le 0.5~\rm{GeV}^2$, the contribution from longitudinally polarized $J/\psi$ LFWFs is negligible. For transverse polarization, the BLFQ LFWFs are dominated by the S-wave component, with minimal contribution from P-wave or higher orbital angular momentum states.
In the small $|\vec{r}_\perp|$ region, although the S-wave amplitude of the BLFQ LFWFs is larger than that of the dipole model LFWFs, ${\cal T}_{gg}$ is close to zero, so the larger S-wave amplitude does not lead to a significantly higher cross section. Conversely, the ``boosted gaussian" LFWFs contribute more through the P-wave than the BLFQ LFWFs. Thus, the replacement of the $J/\psi$ LFWFs by the BLFQ $J/\psi$ wave function also contributes to the observed reduction in the cross section.

Our numerical results can serve as initial conditions for the BK equation~\cite{Balitsky:1995ub,Kovchegov:1999yj} to investigate the proton structure at small $x$ region, which will be explicitly treated in the future study.

\section{Summary\label{Sec6}}
Employing the proton LFWFs obtained using the BLFQ approach, we first evaluate the dipole scattering amplitude in photon-proton diffractive scattering. This amplitude is then combined with the photon LFWFs derived from pQCD and the $J/\psi$ LFWFs from BLFQ to calculate the cross section for exclusive $J/\psi$ production. All wave functions are truncated to the valence-quark Fock sector and renormalized to unity, which allows a controlled and predictive calculation while capturing the dominant nonperturbative features of the proton and charmonium.

We compare our differential cross section with that obtained by Dumitru et al.~\cite{Dumitru:2019qec}, which used the ``harmonic oscillator" model for proton LFWFs, ``boosted gaussian" model for $J/\psi$ LFWFs, and pQCD for photon LFWFs. Our results show slightly lower magnitudes but remarkably similar slopes, indicating that the overall $t$-dependence is largely determined by the spatial structure encoded in the wave functions. The BLFQ framework naturally captures the anisotropic behavior of the dipole scattering amplitude and the dominance of the S-wave component in the $J/\psi$, providing a more microscopic understanding of the underlying scattering dynamics.

Importantly, these predictions are obtained using solely the previously computed wave functions, with no parameter adjustments made to fit the cross-section data. This demonstrates the predictive power of the BLFQ approach in describing hadronic structure and diffractive processes. Furthermore, our results can serve as input for the BK equation to study the proton structure at small Bjorken-$x$, offering theoretical support for future electron-ion collider experiments. Beyond exclusive $J/\psi$ production, this framework can be extended to other vector mesons and hadronic systems, providing a unified and systematically improvable approach to study nonperturbative QCD dynamics in light-front quantization.
\begin{acknowledgments}
We thank Shaoyang Jia for many helpful discussions. 
This work is supported by the National Natural Science Foundation of China under Grant No.~12375143 and No.~12305095, by the Natural Science Foundation of Gansu Province, China, Grant No. 26JRRA111, by the Gansu International Collaboration and Talents Recruitment Base of Particle Physics (2023-2027), by the Senior Scientist Program funded by Gansu Province, Grant No.~25RCKA008.
X. Zhao is supported by Key Research Program of Frontier Sciences, Chinese Academy of Sciences, Grant No.~ZDBS-LY-7020, by the Foundation for Key Talents of Gansu Province, by the Central Funds Guiding the Local Science and Technology Development of Gansu Province, Grant No.~22ZY1QA006, by international partnership program of the Chinese Academy of Sciences, Grant No.~016GJHZ2022103FN, and by the Strategic Priority Research Program of the Chinese Academy of Sciences, Grant No.~XDB34000000.
J. P. Vary is supported by the Department of Energy under Grant No.~DE-SC0023692.  A portion of the computational resources were also provided by Taiyuan Advanced Computing Center.
\end{acknowledgments}

\bibliography{jpsi_photoproduction_Ref.bib}

\end{document}